\documentclass[a4paper,12pt]{article}
\usepackage{amsmath}
\usepackage{amsfonts}
\usepackage{amssymb}
\usepackage{graphicx}
\usepackage{multirow}
\usepackage{jheppub1}
\usepackage{slashed}
\usepackage{hyperref}                 
\hypersetup{                            
             colorlinks = true,
	     linkcolor = blue,
	     linktocpage = true,
             citecolor = blue,
             urlcolor = blue
           }
\def\hhref#1{\href{http://arxiv.org/abs/#1}{#1}}

\newcommand{\stau}{\tilde{\tau}}

\newcommand{\neut}{\tilde{\chi}^0_1}
\newcommand{\nneut}{\tilde{\chi}^0_2}
\newcommand{\nnneut}{\tilde{\chi}^0_3}
\newcommand{\nnnneut}{\tilde{\chi}^0_4}

\newcommand{\charg}{\tilde{\chi}^+_1}
\newcommand{\ccharg}{\tilde{\chi}^+_2}

\def\be{\begin{equation}}
\def\ee{\end{equation}}

\def\bea{\begin{eqnarray}}
\def\eea{\end{eqnarray}}

\title{Reconciling the Muon $g-2$ and Dark Matter Relic Density with the LHC Results in Nonuniversal
Gaugino Mass Models}
\author[a]{Subhendra Mohanty}
\author[a]{Soumya Rao}
\author[b]{D.P. Roy}
\affiliation[a]{Physical Research Laboratory,\\ Ahmedabad 380009, India}
\affiliation[b]{Homi Bhabha Centre for Science Education,\\Tata Institute of Fundamental
Research,\\Mumbai-400088, India.}
\emailAdd{mohanty@prl.res.in, soumya@prl.res.in, dproy1@gmail.com}

\abstract{
	Relatively light electroweak superparticle masses are required to satisfy the bulk
	annihilation region of dark matter relic density and account for the observed
	excess of muon $g-2$, while TeV scale squark and gluino masses are required to
	account for the $125$ GeV Higgs boson mass and the negative SUSY search results
	from 7 TeV LHC in most SUSY models.  These two sets of requirements can 
        be reconciled in a simple
	nonuniversal gaugino mass model, which assumes SUSY breaking via a combination of
	two superfields belonging to the singlet and the 200-plet representations of the
	GUT group SU(5). The model can be probed via squark/gluon search with the present
	and future LHC data. In a more general nonuniversal gaugino mass model the squark
	and gluino masses can be raised to the edge of the discovery limit of 14 TeV LHC
	or beyond.  This model can be probed, however, through the search for electroweak
	pair production of the relatively light sleptons and winos with the 14 TeV LHC
	data in future.}

\begin{document}

\maketitle
\flushbottom

\section{Introduction}

The minimal supergravity or the so called constrained minimal supersymmetric standard
model (CMSSM) has universal gaugino and scalar masses $m_{1/2}$ and $m_0$ at the GUT
scale, along with a universal trilinear coupling parameter $A_0$.  Together with the ratio
of the two Higgs vacuum expectation values ($\tan\beta$) and the sign of the higgsino mass
parameter ($\mu$), one has four and half parameters in this model, while the magnitude of
$\mu$ is determined by the radiative electroweak symmetry breaking condition \cite{1}.  A
large part of the SUSY phenomenology over the years has been based on this model because
of its simplicity and the predictive value.  Here the lightest superparticle (LSP), i.e.
the dark matter, is dominantly a bino over the bulk of the model parameter space.  Since
the bino does not carry any gauge charge, its natural annihilation process is via sfermion
exchange.  And the cosmologically compatible dark matter relic density requires rather
small bino and sfermion masses $\sim 100$ GeV.  This is the so called bulk annihilation
region.  Unfortunately the LEP constraint on the light Higgs boson mass ($m_h>114$ GeV)
practically rules out the bulk annihilation region of the CMSSM parameter space \cite{2}.
The remaining cosmologically compatible dark matter relic density regions of this model
like the stau co-annihilation region, the resonant annihilation region and the focus point
region, all require some amount of fine-tuning between independent SUSY mass parameters.

The reported discovery of Higgs boson at LHC by the ATLAS and CMS experiments \cite{3} at
\begin{equation}
	m_h\simeq 125 \mbox{ GeV}
\end{equation}
have stretched the above mentioned LEP constraint to significantly higher values of $m_0$
and $m_{1/2}$ in the CMSSM parameter space \cite{4}.  It now rules out the lower mass
parts of the stau co-annihilation, the resonant annihilation and the focus point regions.
The remaining parts of these cosmologically compatible dark matter relic density regions
correspond to $m_0\gtrsim 1$ TeV, which imply TeV scale masses of the 1st and 2nd
generation sfermions.  Consequently the CMSSM contribution to the muon $g-2$ is much too
small to explain the anomalous excess observed by the BNL experiment \cite{5}, i.e.
\begin{equation}
	\Delta a_\mu = (28.7\pm 8.0) \times 10^{-10}, \label{g-2} 
\end{equation}
where $a_\mu\equiv (g-2)_\mu/2$ \cite{6}.  The detailed analysis of ref.\cite{4} have also
found very similar results for the nonuniversal Higgs mass models (NUHM) \cite{7}.  More
recently Buchmueller et al \cite{8} have supplemented the constraints on the CMSSM and
NUHM parameter spaces coming from the 125 GeV Higgs boson mass \cite{3}, with those coming
from the latest ATLAS results on direct SUSY search with $5$ fb$^{-1}$ LHC data at 7 TeV
\cite{9}, and the $B_s\to \mu^+\mu^-$ results of ATLAS, CDF, CMS and LHCb experiments
\cite{10} along with the latest direct dark matter detection experiment result of the
XENON100 experiment \cite{11}.  The ATLAS result on direct SUSY search \cite{9} reinforces
the exclusion of the low mass part of the stau co-annihilation region.  The $B_s\to
\mu^+\mu^-$ results \cite{10} are effective in the large $\tan\beta$ ($\gtrsim 30$)
region, where the resonant annihilation region of the SUSY dark matter relic density is
also effective.  It reinforces the exclusion of the low mass part of the latter.  Finally,
the latest XENON100 experiment result \cite{11} enhances the exclusion of the low mass
part of the focus point region \cite{12}.  Thus these experiments strengthen the above
mentioned incompatibility between the SUSY explanations of the observed muon $g-2$ anomaly
\cite{5,6} and dark matter relic density with the $125$ GeV Higgs boson mass result from
LHC \cite{3} in both CMSSM and NUHM.  This has led to a wide perception that there may be
an inherent tension between the two sets of results in any simple SUSY model.  For the
parameter scan in a phenomenological MSSM see e.g. ref.\cite{12a}.

In this work we shall try to reconcile the SUSY explanations of observed muon $g-2$
anomaly \cite{5,6} and dark matter relic density \cite{13} with the $125$ GeV Higgs boson
mass reported from LHC \cite{3} along with the results of ref.\cite{9,10,11} in some
simple and predictive nonuniversal gaugino mass models \cite{14}.  It was shown in
\cite{15} that the most natural SUSY explanation of the observed dark matter relic density
(in terms of fine-tuning) via the bulk annihilation region can be reconciled with the
above mentioned Higgs mass bound from LEP \cite{2} in a set of such nonuniversal gaugino
mass models.  In a recent update of this analysis \cite{16} we have shown that the bulk
annihilation region of dark matter relic density can also be reconciled with the $125$ GeV
Higgs boson mass in these models.  The present work is mainly devoted to the 
analysis of the SUSY
contribution to the muon $g-2$ anomaly \cite{5}, while we continue focus on a Higgs boson
mass of 125 GeV.  We shall see that one of these models can indeed account for the
observed muon $g-2$ anomaly \cite{5}.  We also investigate this issue in a more general
nonuniversal gaugino mass model, where we shall find even a closer agreement with the
observed muon $g-2$ anomaly of eq.(\ref{g-2}) without any conflict with the Higgs mass or
the direct SUSY search results from LHC.  In view of the high precission of the dark
matter relic density data \cite{14} we shall consider solutions lying within 3$\sigma$ of
its central value (see eq (11) below). For muon $g-2$ anomaly, with a relatively large
error bar \cite{5}, we shall consider solutions lying within 2$\sigma$ of the central
value (\ref{g-2}). Finally, for the putative Higgs boson mass of 125 GeV \cite{3}, there
is a spread of  about 3 GeV between the two important decay channels and the two
experiments.  Besides there is a theoretical uncertainty of 2-3 GeV in the prediction of
this mass as discussed below. Therefore we shall consider solutions with the predicted
Higgs mass agreeing with the putative value of 125 GeV within 3 GeV.  Since the
three observables have very different levels of theoretical and experimental 
errors, we shall not attempt to evaluate any overall chi-square for fitting 
these three experimental observables.

In section 2, we summarize the essential ingredients of the model.  In section 3, we
present the results for a specific choice of the nonuniversal gaugino mass model, which
can account for the observed muon $g-2$ anomaly.  Then in section 4, we present the
results for a more general nonuniversal gaugino mass model.  We conclude with a brief
summary of our results in section 5.

\section{Nonuniversality of Gaugino Masses in SU(5) GUT}

The gauge kinetic function responsible for the GUT scale gaugino masses originates from
the vacuum expectation value of the F term of a chiral superfield $\Omega$ responsible for
SUSY breaking,
\begin{equation}
	\frac{\langle F_\Omega\rangle_{ij}}{M_{Planck}}\lambda_i\lambda_j,
	\label{}
\end{equation}
where $\lambda_{1,2,3}$ are the U(1), SU(2), SU(3) gaugino fields - bino, wino and gluino.
Since gauginos belong to the adjoint representation of the GUT group, $\Omega$ and
$F_\Omega$ can belong to any of the irreducible representations appearing in their
symmetric product, i.e.
\begin{equation}
	(24\times 24)_{sym}=1+24+75+200
	\label{}
\end{equation}
for the simplest GUT group SU(5).  Thus, the GUT scale gaugino masses for a given
representation of the SUSY breaking superfield are determined in terms of one parameter as
\begin{equation}
	{M_{1,2,3}}^G=C_{1,2,3}^nm_{1/2}^n
	\label{mgut1}
\end{equation}
where \cite{14}
\begin{equation}
	C_{1,2,3}^{1}=(1,1,1),\; C_{1,2,3}^{24}=(-1,-3,2),\; C_{1,2,3}^{75}=(-5,3,1),\;
	C_{1,2,3}^{200}=(10,2,1).
	\label{cn}
\end{equation}

The CMSSM assumes $\Omega$ to be a singlet leading to universal gaugino masses at the GUT
scale.  On the other hand, any of the nonsinglet representations of $\Omega$ would imply
nonuniversal masses via eqs.(\ref{mgut1}) and (\ref{cn}).  These nonuniversal gaugino mass
models are known to be consistent with the universality of gauge couplings at the GUT
scale \cite{14,17}, with $\alpha_G\simeq 1/25$.  The phenomenology of nonuniversal
gauginos arising from nonsinglet $\Omega$ have been widely studied \cite{17a}.

It was assumed in \cite{15} that SUSY is broken by a combination of a singlet and a
nonsinglet superfields belonging to the $1+24$, $1+75$ or 1+200 representations of SU(5).
Then the GUT scale gaugino masses are given in terms of two mass parameters,
\begin{equation}
	M_{1,2,3}^G=C_{1,2,3}^1m_{1/2}^1+C_{1,2,3}^lm_{1/2}^l\quad\mbox{with}\quad l=24,75
	\mbox{ or } 200.
	\label{mgut2}
\end{equation}
It is evident from the above equation that these NUGM models have an extra
gaugino mass parameter than in the CMSSM.
The corresponding weak scale superparticle and Higgs boson masses are given in terms of
these gaugino masses and the universal scalar mass parameter $m_0$ via the RGE.  It was
shown that in these models one can access the bulk annihilation region of dark matter
relic density, while keeping the light Higgs boson mass above the LEP limit of $114$ GeV
\cite{2}.  In order to understand this, one can equivalently consider the two independent
gaugino mass parameters of eq.(\ref{mgut2}) in any one of these models to be $M_1^G$ and
$M_3^G$.  The corresponding weak scale bino LSP mass is given to a good approximation by
the one-loop RGE,
\begin{equation}
	M_1=\left( \frac{\alpha_1}{\alpha_G} \right)M_1^G\simeq \left(\frac{25}{60}\right)
	M_1^G. 
	\label{rge}
\end{equation}

Thus one can choose a relatively small $M_1^G\sim 200$ GeV along with a small $m_0\sim 80$
GeV to ensure a small weak scale bino mass $M_1\sim 80$ GeV along with right slepton
masses of $\sim 100$ GeV.  Then the annihilation of the bino LSP pair via right slepton
exchange
\begin{equation}
	\chi\chi\stackrel{\tilde{l}_R}{\longrightarrow}\bar{l}l
	\label{annbino}
\end{equation}
gives the desired dark matter relic density \cite{13}.  The other mass parameter $M_3^G$
can then be raised to an appropriate level to raise the Higgs boson mass above the LEP
limit with relatively heavy squarks and gluino.  In our update of ref.\cite{16} the Higgs
boson mass was further raised close to the reported value from LHC \cite{3} with the help
of a large negative $A_0$ term.

  \begin{figure}[h!]
	  \centering
	  \includegraphics[width=0.7\textwidth]{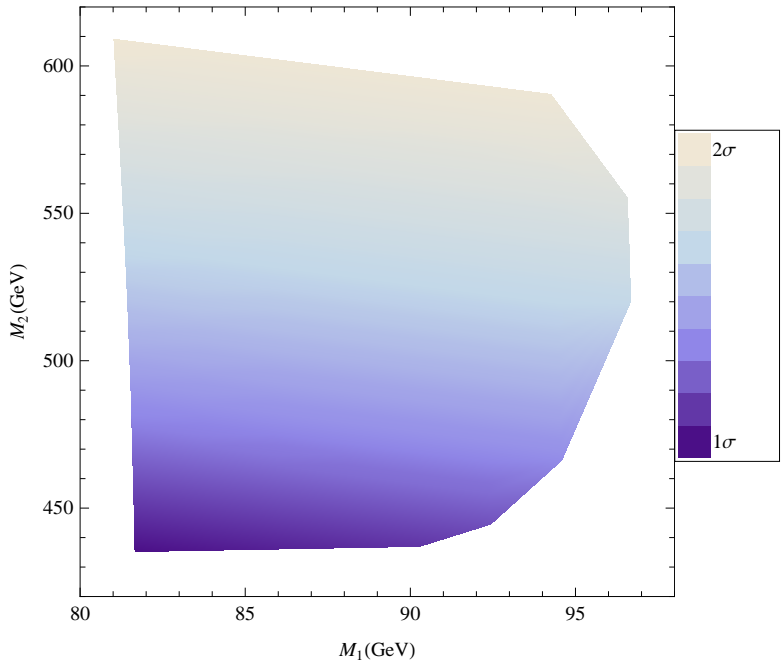}
	  \caption{Parameter space for 1+200 model compatible with the WMAP relic density
	  result and with the predicted $a_\mu$ agreeing with the observed excess $\Delta
	  a_\mu$ (2) within $2\sigma$ in the $M_1-M_2$ plane.  The colour code for
	  $\delta a_\mu= \Delta a_\mu-a_\mu$ is shown on the right.  Here we take
	  $m_0=80$ GeV, $\tan\beta=10$, trilinear couplings $A_{t0}=A_{b_0}=-2.1$ TeV,
	  varying $M_1^G$ between $200-240$ GeV and $M_3^G$ between 600 and 900 GeV.}
	  \label{fig:1}
  \end{figure}

It may be noted here that with given $M_1^G$ and $M_3^G$ inputs, each of the three models
makes a definitive prediction for $M_2^G$.  It can be shown from eqs.(\ref{cn}) and
(\ref{mgut2}) that the $1+200$ model predicts a smaller $M_2^G$ and hence smaller weak
scale wino and left slepton masses compared to the other two models.  Hence it offers the
best chance of accounting for a significant SUSY contribution to the muon $g-2$.
Therefore we shall pursue this issue in detail in the next section using the $1+200$
model.

  \begin{figure}[h!]
	  \centering
	  \includegraphics[width=0.7\textwidth]{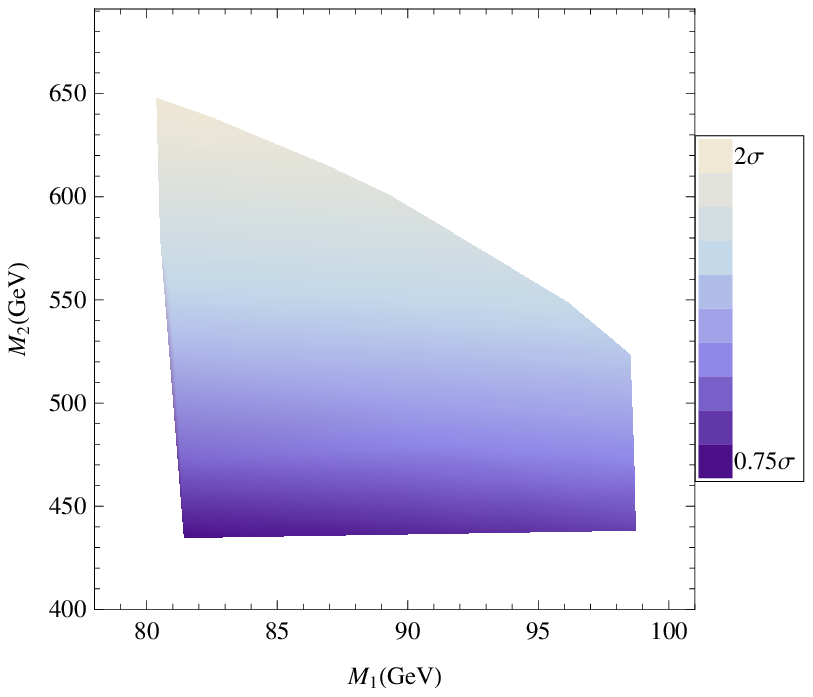}
	  \caption{Same as Fig. 1 , but with $\tan\beta=15$, $m_0=103$ GeV and
	  $A_{t0}=A_{b0}=-1.4$ TeV.}
	  \label{fig:1a}
  \end{figure}

\section[The Weak Scale SUSY spectra and Muon g-2 Prediction of the 1+200 
Model]{The Weak Scale SUSY spectra and Muon $g-2$ Prediction of the $1+200$ 
Model}

\begin{table}[ht!]
    \centering
      All masses in GeV
      \begin{tabular}{|l|l|l|l|l|l|}
	\hline
	\multirow{2}{*}{Particle} & $M_1^G=220$ & \multicolumn{4}{|c|}{$M_1^G=200$}\\
	\cline{3-6}
	& $M_3^G=600$ & $M_3^G=600$ & $M_3^G=700$ & $M_3^G=800$ & $M_3^G=900$ \\
	\hline
	$\neut$ (bino)               & 89.2 & 80.7 & 80.3 & 80.0 & 79.5 \\
	$\nneut$ (wino)              & 447 & 445 & 518 & 591 & 664 \\
	$\nnneut$ (higgsino)         & 1227 & 1226 & 1320 & 1415 & 1511 \\
	$\nnnneut$ (higgsino)        & 1230 & 1230 & 1324 & 1419 & 1515 \\
	$\charg$ (wino)              & 447 & 445 & 518 & 591 & 664 \\
	$\ccharg$ (higgsino)         & 1231 & 1230 & 1324 & 1419 & 1515 \\
	\hline
	$M_1$                   & 90.3 & 81.7 & 81.5 & 81.2 & 80.3 \\
	$M_2$                   & 437 & 435 & 506 & 577 & 648 \\
	$M_3$                   & 1343 & 1343 & 1542 & 1742 & 1933 \\
	$\mu$                        & 1233 &1232 & 1326 & 1421 & 1341 \\
	\hline
	$\tilde{g}$                  & 1354 & 1354 & 1561 & 1766 & 1969 \\
	$\tilde{\tau}_1$             & 101 & 94.8 & 95.3 & 94.8 & 93.5 \\
	$\tilde{\tau}_2$             & 374 & 373 & 424 & 476 & 529 \\
	$\tilde{e}_R,\tilde{\mu}_R$  & 122 & 118 & 117 & 117 & 117 \\
	$\tilde{e}_L,\tilde{\mu}_L$  & 370 & 368 & 421 & 474 & 527 \\
	$\tilde{t}_1$                & 507 & 508 & 729 & 919 & 1094 \\
	$\tilde{t}_2$                & 1049 & 1049 & 1224 & 1395 & 1565 \\
	$\tilde{b}_1$                & 995 & 995 & 1179 & 1358 & 1533 \\
	$\tilde{b}_2$                & 1168 & 1168 & 1345 & 1519 & 1692 \\
	$\tilde{q}_{1,2,R}$          & $\sim 1188$ & $\sim 1188$ & $\sim 1364$ & $\sim
	1538$ & $\sim 1710$  \\
	$\tilde{q}_{1,2,L}$          & $\sim 1234$ & $\sim 1233$ & $\sim 1417$ & $\sim
	1598$ & $\sim 1778$  \\
	$h$			& 122 & 122 & 123 & 123 & 124 \\\hline
	\multicolumn{6}{|c|}{Muon $g-2$} \\\hline
	$a_\mu$ & $1.96\times 10^{-9}$ & $2.04\times 10^{-9}$ &
	$1.66\times 10^{-9}$ & $1.39\times 10^{-9}$ & $1.17\times 10^{-9}$ \\
	$(\delta a_\mu)$ & ($1.14\sigma$) & ($1.04\sigma$) & ($1.51\sigma$) &
	($1.85\sigma$) & ($2.12\sigma$) \\
	\hline
      \end{tabular}
    \caption{The SUSY mass spectrum for the (1+200) model for a $\sim$ 80 GeV LSP and the
	    corresponding $g-2$ contribution from SUSY.  We take $m_0=80$ GeV,
	    $\tan\beta=10$, $A_{t0}=A_{b0}=-2.1$ TeV with $A_{\tau0}=0$ TeV.}
	    \label{tab:spec100}
  \end{table}

We have used the two-loop RGE code \verb|SuSpect| \cite{18} to generate the weak scale
SUSY spectra.  The resulting dark matter relic density and muon anomalous magnetic moment
($g-2$) were computed using the \verb|micrOMEGAs| code \cite{19}.  The sign of the $\mu$
parameter was chosen to be positive for getting a positive SUSY contribution to the muon
anomalous magnetic moment,
\begin{equation}
	a_\mu=\frac{(g-2)_\mu}{2}.
	\label{amu}
\end{equation}
To ensure the bulk annihilation region of the dark matter relic density, we chose a small
$m_0=80$ GeV, and varied $M_1^G$ upwards starting at $200$ GeV.  The second gaugino mass
parameter $M_3^G$ was varied upwards starting from $600$ GeV, to ensure squark and gluino
mass range of interest to LHC.  The $A_0$ parameter was set at $-2.1$ GeV to get the
desired Higgs mass.  We required the dark matter relic density to lie within $3\sigma$
range of WMAP data  \cite{13} i.e.
\begin{equation}
	0.102<\Omega h^2<0.123.
	\label{wmap}
\end{equation}
It effectively limited the $M_1^G$ scan to the $200-240$ GeV range.  Requiring the SUSY
contribution to $a_\mu$ (\ref{amu}) to be within $2\sigma$ of the observed excess of
eq.(\ref{g-2}), restricted the $M_3^G$ scan to the $600-900$ GeV range.

The main SUSY contributions to the muon anomalous magnetic moment $a_\mu$ come from the
bino-right slepton ($\tilde{B}-\tilde{\mu}_R$) and wino-left slepton
($\tilde{W}^0-\tilde{\mu}_L,\tilde{W}^--\tilde{\nu}_L$) loops, which mainly depend on the
weak scale gaugino masses $M_1$ and $M_2$ respectively.  Figure 1 compares the predicted
SUSY contribution to $a_\mu$ with the observed excess $\Delta a_\mu$ of eq(\ref{g-2}), in
the $M_1-M_2$ plane at $\tan\beta=10$, where
\begin{equation}
	\delta a_\mu=\Delta a_\mu-a_\mu.
	\label{damu}
\end{equation}
Figure 2 shows a similar comparison for $\tan\beta=15$ and a larger $m_0$ to compensate
for the lowering of $\tilde{\tau}_1$ at a larger $\tan\beta$.  One sees a little better
agreement in Fig. 2 relative to Fig. 1, resulting from a small rise of the SUSY
contribution with $\tan\beta$.  With futher rise of $\tan\beta$, however, one has to
choose a still larger $m_0$ to compensate for the faster drop of the $\tilde{\tau_1}$
mass.  The resulting increase in the slepton masses compensate the linear rise of $a_\mu$
with $\tan\beta$ at constant SUSY masses.  Thus one gets a broad peak for the predicted
$a_\mu$ at $\tan\beta\simeq 15$. Explicit formulae for the SUSY contributions to $a_\mu$
can be found for example in \cite{Endo:2013bba}.

\begin{table}[ht!]
    \centering
      All masses in GeV
      \begin{tabular}{|l|l|l|l|l|l|}
	\hline
	\multirow{2}{*}{Particle} & $M_1^G=220$ & \multicolumn{4}{|c|}{$M_1^G=200$}\\
	\cline{3-6}
	& $M_3^G=600$ & $M_3^G=600$ & $M_3^G=700$ & $M_3^G=800$ & $M_3^G=900$ \\
	\hline
	$\neut$ (bino)               & 88.8 & 80.2 & 79.9 & 79.6 & 79.1 \\
	$\nneut$ (wino)              & 445 & 443 & 516 & 589 & 662 \\
	$\nnneut$ (higgsino)         & 1032 & 1031 & 1132 & 1233 & 1333 \\
	$\nnnneut$ (higgsino)        & 1036 & 1036 & 1137 & 1237 & 1337 \\
	$\charg$ (wino)              & 445 & 443 & 516 & 589 & 662 \\
	$\ccharg$ (higgsino)         & 1036 & 1037 & 1137 & 1237 & 1337 \\
	\hline
	$M_1$                   & 90.1 & 81.4 & 81.1 & 80.8 & 80.4 \\
	$M_2$                   & 437 & 435 & 506 & 577 & 648 \\
	$M_3$                   & 1331 & 1331 & 1533 & 1734 & 1933 \\
	$\mu$                        & 1034 &1033 & 1133 & 1233 & 1332 \\
	\hline
	$\tilde{g}$                  & 1354 & 1354 & 1561 & 1766 & 1969 \\
	$\tilde{\tau}_1$             & 103 & 97.4 & 96.2 & 93.6 & 89.8 \\
	$\tilde{\tau}_2$             & 381 & 379 & 429 & 480 & 532 \\
	$\tilde{e}_R,\tilde{\mu}_R$  & 138 & 134 & 133 & 133 & 132 \\
	$\tilde{e}_L,\tilde{\mu}_L$  & 374 & 373 & 425 & 477 & 530 \\
	$\tilde{t}_1$                & 748 & 749 & 920 & 1082 & 1240 \\
	$\tilde{t}_2$                & 1107 & 1107 & 1275 & 1441 & 1607 \\
	$\tilde{b}_1$                & 1056 & 1055 & 1232 & 1405 & 1576 \\
	$\tilde{b}_2$                & 1161 & 1161 & 1336 & 1509 & 1681 \\
	$\tilde{q}_{1,2,R}$          & $\sim 1188$ & $\sim 1188$ & $\sim 1364$ & $\sim
	1538$ & $\sim 1710$  \\
	$\tilde{q}_{1,2,L}$          & $\sim 1233$ & $\sim 1233$ & $\sim 1417$ & $\sim
	1598$ & $\sim 1778$  \\
	$h$			& 122 & 122 & 122 & 122 & 123 \\\hline
	\multicolumn{6}{|c|}{Muon $g-2$} \\\hline
	$a_\mu$ & $2.22\times 10^{-9}$ & $2.28\times 10^{-9}$ &
	$1.89\times 10^{-9}$ & $1.59\times 10^{-9}$ & $1.37\times 10^{-9}$ \\
	$(\delta a_\mu)$ & ($0.83\sigma$) & ($0.75\sigma$) & ($1.24\sigma$) &
	($1.61\sigma$) & ($1.89\sigma$) \\
	\hline
      \end{tabular}
      \caption{Same as Table 1, but with $\tan\beta=15$, $m_0=103$ GeV and
	      $A_{t0}=A_{b0}=-1.4$ TeV.}
      \label{tab:spec200}
  \end{table}

For better insight into the underlying physics, we list the weak scale superparticle and
Higgs boson masses along with the resulting $a_\mu$ (\ref{amu}) and $\delta a_\mu$
(\ref{damu}) for $\tan\beta=10$ and $15$ in Tables 1 and 2 respectively.  As in ref.
\cite{16}, the Higgs boson mass has been raised by a few GeV via stop mixing by using a
moderately large and negative GUT scale triliniar coupling parameter $A_0$ for the squark
sector.  The only 
phenomenologically relevant GUT scale $A$ parameters ($A_0$) are $A_{0t} = 
A_{0b}$ and $A_{0\tau}$, where the first two are relevant for the Higgs mass. 
Since in a nonuniversal model the GUT scale $A$ parameter for the lepton sector 
need not be the same as that for quarks, we have kept the $A_\tau = 0$ for 
simplicity.  It should be noted here that the $\overline{\mbox{MS}}$ 
renormalization scheme used in the \verb|SuSpect| RGE code \cite{18} is known 
to predict a lower Higgs boson mass than the on-shell renormalization scheme 
used in FeynHiggs \cite{20} by $2-3$ GeV\cite{21}.  Therefore a
predicted Higgs boson mass $\geq 122$ GeV in these tables is compatible with the reported
mass of $125$ GeV \cite{3} within this theoretical uncertainty.  Coming to the SUSY
masses, one sees that the bino LSP and the right slepton masses are only $\sim 100$ GeV as
expected for the bulk annihilation region.  The wino is at least 5 times heavier than
bino, while the left sleptons are at least $3-4$ times heavier than the right ones.  The
low value of $m_0$ ensures that the left sleptons are always lighter than wino, so that
one expects SUSY cascade decay to result in relatively large LHC signals in the leptonic
channels.  There is also an inverted hierarchy of squark masses suggesting large number of
b-tags in the SUSY signal.  The first two generation squarks are roughly degenerate with
gluinos.  The SUSY search result of the $5$ fb$^{-1}$ data at $7$ TeV in the CMSSM shows a
discovery limit of $1100-1200$ GeV for degenerate squarks and gluons, while the claimed
limit of $1360$ GeV may have questionable physical significance \cite{9}.  With the 20
fb$^{-1}$ data available at 8 TeV one expects this discovery limit to go up to $\sim 1500$
GeV.  If one assumes a similar discovery limit for degenerate squarks and gluinos in the
present model as well, then one would be able to probe the SUSY spectra shown in the first
three columns of tables 1 and 2.  It is evidently imperative to do a dedicated SUSY search
with the accumulated data in this simple nonuniversal gaugino mass model.

\begin{table}[ht!]
    \centering
      All masses in GeV \\
      \begin{tabular}{|l|l|l|l|}
	\hline
	\multirow{2}{*}{Particle} & \multicolumn{3}{|c|}{$M_1^G=200$, $M_2^G=575$ and
	$M_3^G=1200$}\\
	\cline{2-4}
	& $m_0=100$ & $m_0=138$ & $m_0=175$ \\ 
	& $\tan\beta=10$ & $\tan\beta=15$ & $\tan\beta=20$ \\ 
	\hline
	$\neut$ (bino)               & 76.6 & 76.9 & 77.0 \\ 
	$\nneut$ (wino)              & 458 & 459 & 460 \\ 
	$\nnneut$ (higgsino)         & 1666 & 1657 & 1652 \\ 
	$\nnnneut$ (higgsino)        & 1669 & 1659 & 1654 \\ 
	$\charg$ (wino)              & 458 & 459 & 460 \\ 
	$\ccharg$ (higgsino)         & 1669 & 1659 & 1654 \\ 
	\hline
	$M_1$                   & 78.5 & 78.5 & 78.5 \\ 
	$M_2$                   & 442 & 442 & 442 \\ 
	$M_3$                   & 2536 & 2536 & 2537 \\ 
	$\mu$                        & 1663 &1654 & 1649 \\ 
	\hline
	$\tilde{g}$                  & 2580 & 2580 & 2581 \\ 
	$\tilde{\tau}_1$             & 93 & 95.6 & 99.8 \\ 
	$\tilde{\tau}_2$             & 375 & 396 & 420 \\ 
	$\tilde{e}_R,\tilde{\mu}_R$  & 128 & 159 & 192 \\ 
	$\tilde{e}_L,\tilde{\mu}_L$  & 367 & 379 & 394 \\ 
	$\tilde{t}_1$                & 1748 & 1753 & 1757 \\ 
	$\tilde{t}_2$                & 2068 & 2061 & 2053 \\ 
	$\tilde{b}_1$                & 2043 & 2036 & 2026 \\ 
	$\tilde{b}_2$                & 2218 & 2204 & 2186 \\ 
	$\tilde{q}_{1,2,R}$          & $\sim 2231$ & $\sim 2234$ & $\sim 2237$ \\ 
	$\tilde{q}_{1,2,L}$          & $\sim 2249$ & $\sim 2251$ & $\sim 2255$ \\ 
	$h$			& 123 & 123 & 122 \\\hline 
	\multicolumn{4}{|c|}{Muon $g-2$} \\\hline
	$a_\mu$ & $2.47\times 10^{-9}$ & $2.67\times 10^{-9}$ &
	$2.62\times 10^{-9}$ \\ 
	$(\delta a_\mu)$& ($0.51\sigma$) & ($0.26\sigma$) & ($0.32\sigma$) \\ 
	\hline
      \end{tabular}
    \caption{The SUSY mass spectrum for a general non-universal model for a $\sim 80$ GeV
	    LSP with increasing $\tan\beta$ and the corresponding $g-2$ contribution from
	    SUSY.  We take $A_{t0}=A_{b0}=-1.4$ TeV with $A_{\tau 0}=0$, while $m_0$ is
	    chosen to ensure  the correct relic density  in each case.}
	    \label{tab:spec300}
  \end{table}

In closing this section it should be noted that our results are immune to the $B_s\to
\mu^+\mu^-$ constraints \cite{10}, which are effective only in the large $\tan\beta$
($\gtrsim 30$) region.  They are also immune to the direct detection limit from the XENON
100 experiment \cite{11}, since the predicted cross-section is very small for a bino
dominated dark matter.  Detailed account of this comparison is given in ref.\cite{16}.
And finally for the preferred sign of $\mu$, we find that the branching fraction
$\mathcal{B}(b\to s\gamma)$ for our chosen benchmark points falls within $2\sigma$ of the
experimental world average $\mathcal{B}(b\to s\gamma)=(3.55\pm 0.25)\times 10^{-4}$
\cite{16a}.

\section[The Weak SUSY spectra and Muon g-2 Prediction of a General 
Nonuniversal Gaugino Mass Model]{The Weak SUSY spectra and Muon $g-2$ 
Prediction of a General Nonuniversal Gaugino Mass Model}

Finally we shall extend the above analysis to a general nonuniversal gaugino mass model,
where all the three GUT scale gaugino masses $M_1^G$, $M_2^G$ and $M_3^G$ are independent
parameters.  This means that there are now two more gaugino mass parameters 
than in the CMSSM.  This model can be realized in a scenario of SUSY breaking 
by three superfields, belonging to different adjoint representations of the GUT 
group e.g. a ($1+75+200$) model.  One can equivalently choose the $M_1^G$, 
$M_2^G$ and $M_3^G$ as the three input parameters.

As in the previous section the $M_1^G$ and $m_0$ parameters are chosen to ensure adherence
to the bulk annihilation region of dark matter relic density.  The $M_2^G$ parameter can
now be chosen to obtain a SUSY contribution to $a_\mu$ very close to the observed excess
of eq.(\ref{g-2}).  Then the remaining parameter $M_3^G$ can be chosen to be in the TeV
scale so that one can account for the reported Higgs mass of 125 GeV \cite{3} with an
$A_0$ parameter of similar size as $M_3^G$.

\begin{table}[t!]
    \centering
      All masses in GeV\\
      \begin{tabular}{|l|l|l|l|}
	\hline
	\multirow{2}{*}{Particle} & \multicolumn{3}{|c|}{$M_1^G=200$, $M_2^G=575$ and
	$M_3^G=1500$}\\
	\cline{2-4}
	& $m_0=117$ & $m_0=160$ & $m_0=201$ \\ 
	& $\tan\beta=10$ & $\tan\beta=15$ & $\tan\beta=20$ \\ 
	\hline
	$\neut$ (bino)               & 74.8 & 75.0 & 75.2 \\ 
	$\nneut$ (wino)              & 455 & 456 & 457 \\ 
	$\nnneut$ (higgsino)         & 1976 & 1965 & 1959 \\ 
	$\nnnneut$ (higgsino)        & 1977 & 1967 & 1961 \\ 
	$\charg$ (wino)              & 455 & 456 & 457 \\ 
	$\ccharg$ (higgsino)         & 1978 & 1967 & 1961 \\ 
	\hline
	$M_1$                   & 76.7 & 76.7 & 76.8 \\ 
	$M_2$                   & 437 & 437 & 438 \\ 
	$M_3$                   & 3125 & 3126 & 3128 \\ 
	$\mu$                        & 1970 & 1960 & 1954 \\ 
	\hline
	$\tilde{g}$                  & 3178 & 3179 & 3180 \\ 
	$\tilde{\tau}_1$             & 93.9 & 97.7 & 104 \\ 
	$\tilde{\tau}_2$             & 375 & 402 & 434 \\ 
	$\tilde{e}_R,\tilde{\mu}_R$  & 139 & 176 & 214 \\ 
	$\tilde{e}_L,\tilde{\mu}_L$  & 363 & 379 & 398 \\ 
	$\tilde{t}_1$                & 2207 & 2212 & 2216 \\ 
	$\tilde{t}_2$                & 2535 & 2529 & 2520 \\ 
	$\tilde{b}_1$                & 2517 & 2510 & 2499 \\ 
	$\tilde{b}_2$                & 2724 & 2709 & 2689 \\ 
	$\tilde{q}_{1,2,R}$          & $\sim 2738$ & $\sim 2741$ & $\sim 2745$ \\ 
	$\tilde{q}_{1,2,L}$          & $\sim 2748$ & $\sim 2751$ & $\sim 2754$ \\ 
	$h$			& 123 & 123 & 122 \\\hline 
	\multicolumn{4}{|c|}{Muon $g-2$} \\\hline
	$a_\mu$ & $2.67\times 10^{-9}$ & $2.7\times 10^{-9}$ &
	$2.54\times 10^{-9}$ \\ 
	$(\delta a_\mu)$& ($0.26\sigma$) & ($0.23\sigma$) & ($0.43\sigma$) \\ 
	\hline
      \end{tabular}
      \caption{The SUSY mass spectrum for a general non-universal model for a $\sim 80$
	    GeV LSP with increasing $\tan\beta$ and the corresponding $g-2$ contribution
	    from SUSY.  We take $A_{t0}=A_{b0}=-1.4$ TeV with $A_{\tau 0}=0$ TeV as
	    before, while  $m_0$ is chosen such that the correct relic density is obtained
	    in each case.}
	    \label{tab:spec400}
  \end{table}

Table 3 lists the weak scale superparticle and Higgs boson masses for such a model along
with resulting $a_\mu$ predictions for $\tan\beta=10,15$ and 20.  The $M_1^G$ value is
chosen as in the last section to ensure adherence to the bulk annihilation region.  In
this case one requires a somewhat larger value of $m_0$ to ensure $\tilde{\tau}_1$ mass to
remain $\sim 20\%$ above the bino LSP mass to avoid copious co-annihilation, so that the
dark matter relic density remains in the desired range of eq.(\ref{wmap}).  Note that the
$m_0$ value goes up with $\tan\beta$ to compensate for the decrease of $\stau_1$ mass from
RGE, as mentioned in the last section.  The value of $M_2^G=575$ GeV is chosen to account
for the observed $a_\mu$ excess to within a quarter $\sigma$.  The value of $M_3^G=1200$
GeV is chosen to account for the reported Higgs boson mass \cite{3} with a similar size of
$A_0=-1400$ GeV.  The resulting degenerate squark-gluino masses are in the range of
$2300-2600$ GeV, which can be probed only by the 14 TeV LHC experiments.  Finally table 4
shows the corresponding weak scale superparticle and Higgs boson masses along with the
$a_\mu$ predictions for a higher $M_3^G=1500$ GeV.  The results are very similar to those
of table 3, except for rise of the degenerate squark-gluino masses to the range of
2800-3200 GeV.  This may be at the edge of the discovery limit of 14 TeV LHC run.  The
squark-gluino masses can be pushed up still higher with a higher value of $M_3^G$.  Note
however, that one can search for the pair production of relatively light sleptons and also
winos via the Drell-Yan process with the 14 TeV LHC data.  Indeed this provides a direct
LHC test for the SUSY explanation of the observed excess of $a_\mu$ (\ref{g-2}) via
relatively low mass sleptons and wino.  The discovery limit with 5 fb$^{-1}$ of 7 TeV LHC
data by the ATLAS collaboration \cite{22} falls short of the slepton and wino mass ranges
of our interest.  It is imperative to make a dedicated search for the electroweak
production of such superparticle pairs in this model with the available LHC data, of about
5 and 20 fb$^{-1}$ at 7 and 8 TeV respectively.

\section{Conclusion}

The relatively low SUSY masses favoured by the observed dark matter relic density
\cite{13} and especially the observed excess of muon $g-2$ \cite{5} are incompatible with
the reported Higgs boson mass of 125 GeV \cite{3} and the direct SUSY search results
\cite{9} from 7 TeV LHC in the CMSSM as well as the NUHM \cite{4,8}.  However, these two
sets of results can be reconciled in a simple and predictive nonuniversal gaugino mass
model, based on the SUSY GUT group SU(5) \cite{14}.  It assumes SUSY breaking by a
combination of a singlet and a non singlet superfields belonging to the symmetric product
of two adjoint representations of the GUT group, i.e. $1+24, 1+75$ or $1+200$
representations \cite{15}.  In each case one can satisfy the bulk annihilation region of
dark matter relic density with relatively small bino and right slepton masses $\sim 100$
GeV, while having TeV scale squark/gluino masses and $A_0$ parameter to satisfy the Higgs
mass and direct SUSY search results from 7 TeV LHC \cite{16}.  We show here that the 1+200
model predicts a relatively modest mass range for wino and left slepton masses, which can
also account for the observed excess of the muon $g-2$.  Part of this model parameter
space can be probed via squark/gluino search with the available LHC data, while the
remainder can be probed with the 14 TeV LHC data.  We then present a more general model of
nonuniversal gaugino masses, where one can account for the bulk annihilation region of
dark matter relic density and the observed excess of muon $g-2$, while pushing up the
squark/gluino masses beyond the reach of the available 7 and 8 TeV LHC data and in fact to
the edge of the discovery limit of the 14 TeV data or beyond.  However, the model can be
probed via SUSY search for electroweak production of the relatively light wino and slepton
pairs at least with the 14 TeV LHC data.  We conclude with the hope that the ATLAS and CMS
collaborations will start dedicated search for wino and sleptons in these simple models
via squark/gluino cascade decay as well as electroweak pair production.

\section{Acknowledgement}

The work of DPR was partly supported by the senior scientist fellowship of
Indian National Science Academy.


\begin{thebibliography}{99}

\bibitem{1} See e.g. ``Perspectives in Supersymmetry '', G.~L.~Kane, (ed.), World
		Scientific (1998); M.~Drees, R.~Godbole and P.~Roy,
``Theory and phenomenology of sparticles: An account of four-dimensional N=1 supersymmetry in high energy physics,''
  Hackensack, USA: World Scientific (2004); H.~Baer and X.~Tata,
``Weak scale supersymmetry: From superfields to scattering events,''
  Cambridge, UK: Univ. Pr. (2006).

\bibitem{2} K.~Nakamura {\it et al.}  [Particle Data Group Collaboration], J.\ Phys.\ G {\bf 37}, 075021 (2010).

\bibitem{3} G.~Aad {\it et al.}  [ATLAS Collaboration],
  Phys.\ Lett.\ B {\bf 716}, 1 (2012)
  [arXiv:\hhref{1207.7214} [hep-ex]]; S.~Chatrchyan {\it et al.}  [CMS Collaboration],
  Phys.\ Lett.\ B {\bf 716}, 30 (2012)
  [arXiv:\hhref{1207.7235} [hep-ex]].

  \bibitem{4} O.~Buchmueller, R.~Cavanaugh, A.~De Roeck, M.~J.~Dolan, J.~R.~Ellis, H.~Flacher, S.~Heinemeyer and G.~Isidori {\it et al.},
  Eur.\ Phys.\ J.\ C {\bf 72}, 2020 (2012)
  [arXiv:\hhref{1112.3564} [hep-ph]]; H.~Baer, V.~Barger and A.~Mustafayev,
  Phys.\ Rev.\ D {\bf 85}, 075010 (2012)
  [arXiv:\hhref{1112.3017} [hep-ph]];
  J.~Cao, Z.~Heng, D.~Li and J.~M.~Yang,
  Phys.\ Lett.\ B {\bf 710}, 665 (2012)
  [arXiv:\hhref{1112.4391} [hep-ph]];  A.~Fowlie, M.~Kazana, K.~Kowalska, S.~Munir, L.~Roszkowski, E.~M.~Sessolo, S.~Trojanowski and Y.~-L.~S.~Tsai,
  Phys.\ Rev.\ D {\bf 86}, 075010 (2012)
  [arXiv:\hhref{1206.0264} [hep-ph]].

  \bibitem{5} G.~W.~Bennett {\it et al.}  [Muon (g-2) Collaboration],
  Phys.\ Rev.\ D {\bf 80}, 052008 (2009)
  [arXiv:\hhref{0811.1207} [hep-ex]].

  \bibitem{6} M.~Davier, A.~Hoecker, B.~Malaescu and Z.~Zhang,
  Eur.\ Phys.\ J.\ C {\bf 71}, 1515 (2011)
  [Erratum-ibid.\ C {\bf 72}, 1874 (2012)]
  [arXiv:\hhref{1010.4180} [hep-ph]].

  \bibitem{7} H.~Baer, A.~Mustafayev, S.~Profumo, A.~Belyaev and X.~Tata,
  JHEP {\bf 0507}, 065 (2005)
  [\hhref{hep-ph/0504001}]; H.~Baer, A.~Mustafayev, S.~Profumo, A.~Belyaev and X.~Tata,
  Phys.\ Rev.\ D {\bf 71}, 095008 (2005)
  [\hhref{hep-ph/0412059}]; J.~R.~Ellis, K.~A.~Olive and P.~Sandick,
  Phys.\ Rev.\ D {\bf 78}, 075012 (2008)
  [arXiv:\hhref{0805.2343} [hep-ph]].

  \bibitem{8} O.~Buchmueller, R.~Cavanaugh, M.~Citron, A.~De Roeck, M.~J.~Dolan, J.~R.~Ellis, H.~Flacher and S.~Heinemeyer {\it et al.},
  Eur.\ Phys.\ J.\ C {\bf 72}, 2243 (2012)
  [arXiv:\hhref{1207.7315} [hep-ph]].

  \bibitem{9} V.~A.~Mitsou [ATLAS Collaboration],
	  arXiv:\hhref{1210.1679} [hep-ex];
  G.~Aad {\it et al.}  [ATLAS Collaboration],
  Eur.\ Phys.\ J.\ C {\bf 73}, 2362 (2013)
  [arXiv:\hhref{1212.6149} [hep-ex]].

  \bibitem{10} G.~Aad {\it et al.}  [ATLAS Collaboration],
  Phys.\ Lett.\ B {\bf 713}, 387 (2012)
  [arXiv:\hhref{1204.0735} [hep-ex]]; T.~Aaltonen {\it et al.}  [CDF Collaboration],
  Phys.\ Rev.\ Lett.\  {\bf 107}, 191801 (2011)
  [arXiv:\hhref{1107.2304} [hep-ex]].

  \bibitem{11} E.~Aprile {\it et al.}  [XENON100 Collaboration],
  Phys.\ Rev.\ Lett.\  {\bf 109}, 181301 (2012)
  [arXiv:\hhref{1207.5988} [astro-ph.CO]]; E.~Aprile {\it et al.}  [XENON100 Collaboration],
  Phys.\ Rev.\ Lett.\  {\bf 107}, 131302 (2011)
  [arXiv:\hhref{1104.2549} [astro-ph.CO]].

  \bibitem{12} J.~L.~Feng, K.~T.~Matchev and D.~Sanford,
  Phys.\ Rev.\ D {\bf 85}, 075007 (2012)
  [arXiv:\hhref{1112.3021} [hep-ph]].

  \bibitem{12a} T.~Han, Z.~Liu and A.~Natarajan,
	  arXiv:\hhref{1303.3040} [hep-ph].

  \bibitem{13} E.~Komatsu {\it et al.}  [WMAP Collaboration],
  Astrophys.\ J.\ Suppl.\  {\bf 180}, 330 (2009)
  [arXiv:\hhref{0803.0547} [astro-ph]].

  \bibitem{14} J.~R.~Ellis, K.~Enqvist, D.~V.~Nanopoulos and K.~Tamvakis,
  Phys.\ Lett.\ B {\bf 155}, 381 (1985); M.~Drees,
  Phys.\ Lett.\ B {\bf 158}, 409 (1985).

  \bibitem{15} S.~F.~King, J.~P.~Roberts and D.~P.~Roy,
  JHEP {\bf 0710}, 106 (2007)
  [arXiv:\hhref{0705.4219} [hep-ph]].

  \bibitem{16} S.~Mohanty, S.~Rao and D.~P.~Roy,
  JHEP {\bf 1211}, 175 (2012)
  [arXiv:\hhref{1208.0894} [hep-ph]].

  \bibitem{16a} 
	    Y.~Amhis {\it et al.}  [Heavy Flavor Averaging Group Collaboration],
	      arXiv:1207.1158 [hep-ex].

  \bibitem{17} U.~Chattopadhyay and P.~Nath,
  Phys.\ Rev.\ D {\bf 65}, 075009 (2002)
  [\hhref{hep-ph/0110341}].

  \bibitem{17a} G.~Anderson, H.~Baer, C.~-h.~Chen and X.~Tata,
  Phys.\ Rev.\ D {\bf 61}, 095005 (2000)
  [\hhref{hep-ph/9903370}]; K.~Huitu, Y.~Kawamura, T.~Kobayashi and K.~Puolamaki,
  Phys.\ Rev.\ D {\bf 61}, 035001 (2000)
  [\hhref{hep-ph/9903528}]; U.~Chattopadhyay and D.~P.~Roy,
  Phys.\ Rev.\ D {\bf 68}, 033010 (2003)
  [\hhref{hep-ph/0304108}].

  \bibitem{18} A.~Djouadi, J.~-L.~Kneur and G.~Moultaka,
  Comput.\ Phys.\ Commun.\  {\bf 176}, 426 (2007)
  [\hhref{hep-ph/0211331}].

  \bibitem{19} G.~Belanger, F.~Boudjema, P.~Brun, A.~Pukhov, S.~Rosier-Lees, P.~Salati and A.~Semenov,
  Comput.\ Phys.\ Commun.\  {\bf 182}, 842 (2011)
  [arXiv:\hhref{1004.1092} [hep-ph]]; G.~Belanger, F.~Boudjema, A.~Pukhov and A.~Semenov,
  Comput.\ Phys.\ Commun.\  {\bf 174}, 577 (2006)
  [\hhref{hep-ph/0405253}]; G.~Belanger, F.~Boudjema, A.~Pukhov and A.~Semenov,
  Comput.\ Phys.\ Commun.\  {\bf 149}, 103 (2002)
  [\hhref{hep-ph/0112278}].

\bibitem{Endo:2013bba}
  T.~Moroi,
  Phys.\ Rev.\ D {\bf 53}, 6565 (1996)
  [Erratum-ibid.\ D {\bf 56}, 4424 (1997)]
  [\hhref{hep-ph/9512396}];
  M.~Endo, K.~Hamaguchi, S.~Iwamoto and T.~Yoshinaga,
  arXiv:\hhref{1303.4256} [hep-ph].  

  \bibitem{20} T.~Hahn, S.~Heinemeyer, W.~Hollik, H.~Rzehak and G.~Weiglein,
  Comput.\ Phys.\ Commun.\  {\bf 180}, 1426 (2009).

  \bibitem{21} A.~Arbey, M.~Battaglia, A.~Djouadi and F.~Mahmoudi,
arXiv:[\hhref{1207.1348} [hep-ph]; S.~Heinemeyer, O.~Stal and G.~Weiglein,
Phys.\ Lett.\ B {\bf 710}, 201 (2012)
[arXiv:\hhref{1112.3026} [hep-ph]]; B.~C.~Allanach, A.~Djouadi, J.~L.~Kneur, W.~Porod and P.~Slavich,
JHEP {\bf 0409}, 044 (2004)
[\hhref{hep-ph/0406166}]; G.~Degrassi, S.~Heinemeyer, W.~Hollik, P.~Slavich and G.~Weiglein,
Eur.\ Phys.\ J.\ C {\bf 28}, 133 (2003)
[\hhref{hep-ph/0212020}]; R.~V.~Harlander, P.~Kant, L.~Mihaila and M.~Steinhauser,
Phys.\ Rev.\ Lett.\ {\bf 100}, 191602 (2008)
[Phys.\ Rev.\ Lett.\ {\bf 101}, 039901 (2008)]
[arXiv:\hhref{0803.0672} [hep-ph]]; S.~P.~Martin,
  Phys.\ Rev.\ D {\bf 75}, 055005 (2007)
  [\hhref{hep-ph/0701051}].

  \bibitem{22} G.~Aad {\it et al.}  [ATLAS Collaboration],
  Phys.\ Lett.\ B {\bf 718}, 879 (2013)
  [arXiv:\hhref{1208.2884} [hep-ex]].
\end{thebibliography}
\end{document}